\documentclass{article}

\PassOptionsToPackage{numbers, compress}{natbib}

\usepackage[preprint]{neurips_2025}

\usepackage[utf8]{inputenc} 
\usepackage[T1]{fontenc}    
\usepackage{hyperref}       
\usepackage{url}            
\usepackage{booktabs}       
\usepackage{amsfonts}       
\usepackage{nicefrac}       
\usepackage{microtype}      
\usepackage{xcolor}         

\usepackage{graphicx}
\usepackage{amssymb}
\usepackage{amsmath}
\usepackage{cases}
\usepackage{algorithm}
\usepackage{subcaption}
\usepackage{algpseudocode}
\usepackage{multirow}
\usepackage{wrapfig} 
\usepackage{multicol} 

\algnewcommand{\algorithmicinput}{\textbf{Input:}}
\algnewcommand{\Input}{\item[\algorithmicinput]}
\algnewcommand{\algorithmicoutput}{\textbf{Output:}}
\algnewcommand{\Output}{\item[\algorithmicoutput]}
\algnewcommand{\algorithmicparameters}{\textbf{Parameters:}}
\algnewcommand{\Parameters}{\item[\algorithmicparameters]}
\algrenewcommand{\algorithmiccomment}[1]{\hfill\(\triangleright\) #1}

\newcommand{\floor}[1]{\left\lfloor #1 \right\rfloor}

\title{MNO : A Multi-modal Neural Operator for Parametric Nonlinear BVPs}

\author{%
  Vamshi C.~Madala \\
  Department of Electrical and Computer Engineering\\
  University of California Santa Barbara\\
  \texttt{vamshichowdary@ucsb.edu} \\
  \And
  Nithin Govindarajan \\
  Department of Electrical Engineering\\
  KU Leuven\\
  \texttt{nithin.govindarajan@kuleuven.be} \\
  \And
  Shivkumar Chandrasekaran \\
  Department of Electrical and Computer Engineering\\
  University of California Santa Barbara\\
  \texttt{shiv@ucsb.edu} \\
}

\begin{document}

\maketitle

\begin{abstract}
We introduce a novel Multi-modal Neural Operator (MNO) architecture designed to learn solution operators for multi-parameter non-linear boundary value problems (BVPs). Traditional neural operators primarily map either the PDE coefficients or source terms independently to the solution, limiting their flexibility and applicability. In contrast, our proposed MNO architecture generalizes these approaches by mapping multiple parameters—including PDE coefficients, source terms, and boundary conditions—to the solution space in a unified manner. 
Our MNO is motivated by the hierarchical nested bases of the Fast Multipole Method (FMM) and is constructed systematically through three key components: a parameter-efficient Generalized FMM (GFMM) block, a Uni-modal Neural Operator (UNO)  built upon GFMM-blocks for single-parameter mappings, and most importantly, a multi-modal fusion mechanism extending these components to learn the joint map. 
We demonstrate the multi-modal generalization capacity of our approach on both linear and nonlinear BVPs. Our experiments show that the network effectively handles simultaneous variations in PDE coefficients and source/boundary terms\footnote{github.com/vamshichowdary/MultimodalNeuralOperator}.
\end{abstract}

\section{Introduction}
We are interested in simultaneously solving a family of variable-coefficient partial differential equations (PDEs) with inhomogeneous boundary conditions on a open connected domain $\Omega$ for $u: \bar{\Omega} \to \mathbb{R}$. Specifically, our aim is to solve
\begin{subequations} 
\label{eqn:gen_pde} 
\begin{eqnarray}
    \mathcal{D}_1 (x,u; a) = c(x), & x \in \Omega,\\ 
    \mathcal{D}_2(x,u; f) = g(x), & x \in \partial\Omega,
\end{eqnarray}
\end{subequations}
where $\mathcal{D}_1$ is a nonlinear differential operator and $\mathcal{D}_2$ another (not necessarily) differential operator of lower degree. Furthermore the coefficients $a,c,f,g$ will be assumed to be in some compact set during training.

\paragraph{Related works} In recent years, data-driven neural network-based approaches to solve PDEs have made considerable progress, of which, physics-informed neural networks (PINNs)\citep{raissi2019physics} and neural operators (NO)~\citep{kovachki2023neural,deeponet2021} are the most popular. PINNs originally took a generic DNN architecture and found the solution by setting the residual of the PDE as the loss function. They have evolved since then to be versatile and can exploit the GPU easily for solving complex nonlinear PDEs. But they can take an unpredictable amount of time to reduce the residual to a desired level, becoming very slow in some circumstances. NOs on the other hand, are pre-trained over a compact family of coefficients to produce the correct solution. So their training time can be much larger than that of a PINN, but their inference time is fixed and typically quite fast. Training NOs requires knowledge of true solutions, whereas PINNs don't have that constraint. Our setup is akin to NOs rather than PINNs.

Many variations of neural operators have been proposed in recent years to tackle different aspects of PDEs such as multiple scales, irregular grids\citep{Li2020MultipoleGNA, hao2023gnot, beno2024}, operator domains\citep{fno2021, Helwig2023GroupEFA, Gupta2021MultiwaveletbasedOLA, nsno2024} and underlying neural network architectures\citep{Raonic2023ConvolutionalNOA, Wu2024TransolverAFA}. Neural operators have been applied in a variety of domains like computational fluid dynamics\citep{mao2021deepm}, multi-physics\citep{cai2021deepm}, wave propagation\citep{nsno2024}, weather forecasting\citep{pathak2022fourcastnet}, etc. Nevertheless, most existing neural operators either map only the operator coefficient $a$ or the source term $c$ of the PDE to the solution $u$, but not both simultaneously. For example,~\citet{nsno2024} used a combined FNO and U-Net~\citep{ronneberger2015u} based architecture to decouple the coefficient and source term in the Helmholtz equation to learn a joint map. Moreover, most neural operator approaches assume fixed boundary conditions~\citep{aldirany2024multi, sau2024reviewneuralnetworksolvers} or propose complex extensions to handle inhomogeneous boundary conditions such as BENO~\citep{beno2024} where the authors use separate graph neural networks for interior and boundary source terms through connecting them with a transformer\citep{vaswani2017attention} network.

We refer to the operators that map only $a \mapsto u$ or $c \mapsto u$ as \emph{uni-modal} operators, and operators that map both i.e.,  $a,c \mapsto u$, or more, as \emph{multi-modal} operators. The lack of truly multi-modal neural operators in the literature partly stems from the mismatch between the types of objects being mapped: $a,f$ parametrizes a family of operators that can act on $u$, whereas $c,g$ parametrizes the range space of these operators. They typically correspond to different physical quantities with different units. The design of models that effectively handle both relations is highly useful in engineering applications. In inverse scattering, the Helmholtz equation $\Delta u + k^2(1+a(x))u(x) = c(x)$ is routinely solved multiple times for different scattering potentials $a$ and excitations $c$~\citep{nsno2024}. In structural mechanics, elasticity problems require repeated solving of PDEs to retrieve displacement fields $u$ for varying material properties $a$ and load configurations $c$. Thus, a multi-modal operator that can generalize across both material and forcing variations could reduce simulation costs and enable real-time feedback during design iterations.

\paragraph{Our contributions} 
In this work, we propose a novel \textit{multi-modal neural operator} (MNO) architecture that learns solution operators mapping multiple parameters of PDEs, boundary equations and their right hand sides (RHS) to the solution. That is, our network learns the following general solution operator for (\ref{eqn:gen_pde}):
\begin{eqnarray*}
    \mathcal{A}^\dagger : a,c,f,g \mapsto u.
\end{eqnarray*}
Our proposed architecture is rigorously motivated by the hierarchical nested bases of the fast multipole method (FMM)\citep{greengard1987fast}. While previous works~\citep{fan2019multiscale, sushnikova2022fmm, Li2020MultipoleGNA} have proposed FMM-based architectures for unimodal operators that map either $a$ or $c$ to $u$ for fixed boundary conditions, ours is the first to extend this architecture to the multi-modal paradigm where all the coefficients of the PDE are inputs of the network. To achieve this functionality, our MNO is systematically built up in increasing levels of complexity as follows:
\begin{itemize}
    \item \textbf{GFMM-block}: a parameter-efficient feed-forward network that generalizes the FMM signal flow graph for multiple parameter inputs with nonlinear activations, referred to as the Generalized FMM (GFMM) block (section \ref{sec:fmm-block}).
    \item \textbf{UNO}: a Uni-modal Neural Operator (UNO) built on top of Generalized FMM-blocks for learning solution operators of PDEs with one parameter (e.g., $a$ or $c$) as input (section \ref{sec:uno}).
    \item \textbf{Multi-modal Fusion}: a \textit{multi-linear} extension of the UNO and GFMM-blocks called multi-modal fusion to learn the joint map $a, c \mapsto u$ (section \ref{sec: multi_fusion}). 
\end{itemize}
We evaluate the performance of our approach on both linear and nonlinear boundary value problems (BVPs), considering both uni-modal and multi-modal scenarios. For the linear case, we use Darcy’s flow, which simplifies to the Poisson equation under constant parameters. For the nonlinear case, we test our method on a first-order BVP featuring an absolute value nonlinearity. While we compare the performance of our UNO model against existing architectures such as FNO and DeepONet—showing improved generalization—we focus primarily on demonstrating the effectiveness of our MNO architecture. In particular, we show that MNO can achieve multi-modal generalization without a significant performance drop compared to its uni-modal (UNO) counterpart.

In our experiments, we put explicit effort into testing the generalization ability of our network to out-of-distribution data. Training neural operators requires ground-truth solutions for different input parameter functions.  In most cases, training data is generated by constructing examples using a classical solver~\citep{deeponet2021, fno2021, long2018pde, wang2021learning} which might be prohibitively expensive, especially for high-dimensional and nonlinear PDEs\citep{gomez2024beginner}. Given that it is much easier to sample the solution directly by using the PDE's differential operator to compute the RHS~\citep{hasani2024generating}, there is much scope for training the neural operators on this synthetic data and employ them on unseen parameter distributions and PDEs~\citep{hasani2024generating, chen2024data, totounferoush2025paving, shen2024ups}. The residuals of the predicted solution when trained using this synthetic data, however, can exhibit a different distribution compared to the actual distributions observed during testing~\citep{lerer2024multigrid}. This discrepancy has an impact on the generalization capacity of the network.

\vspace{-0.2cm}
\section{The multi-modal neural operator (MNO)}
\begin{figure}
    \centering
    \includegraphics[width=0.7\textwidth]{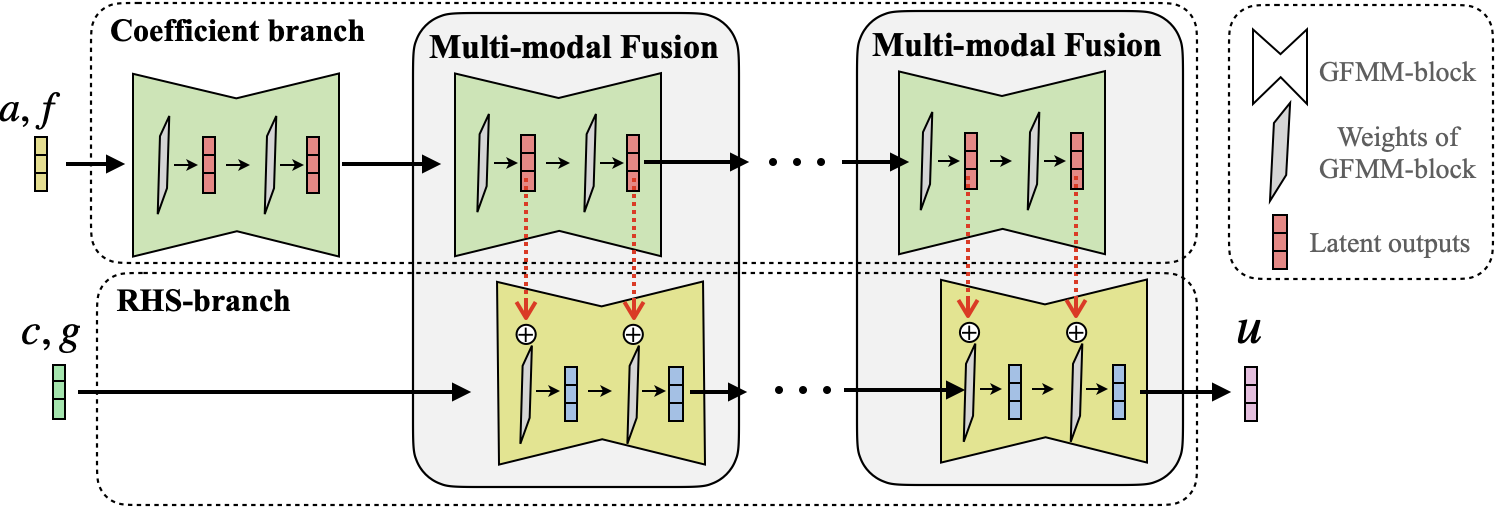}
    \caption{\small {A schematic overview of the multi-modal neural operator (MNO).}}
    \label{fig:mno_overview}
\end{figure}
The overview of our multi-modal neural operator (MNO) is shown in Figure \ref{fig:mno_overview}. It consists of multiple GFMM-blocks stacked along two branches. The top branch (coefficient-branch) accepts $a,f$ as inputs and the bottom branch (RHS-branch) takes in $c,g$ as inputs. The hierarchical nested structure of each GFMM-block produces intermediate basis representations of the respective PDE parameter inputs, shown as \emph{latent outputs} in the figure and described in detail in the next section. In section \ref{sec:uno}, we develop the uni-modal neural operator (UNO) that learns the map $c \mapsto u$ for fixed coefficient PDEs. Finally, in section \ref{sec: multi_fusion}, we present the multi-fusion operation in MNO that updates the RHS-branch by learning the weight corrections from the latent outputs of the coefficient branch for variable coefficient PDEs.

\vspace{-0.2cm}
\subsection{The GFMM-block} \label{sec:fmm-block}

\begin{figure}
    \centering
    \includegraphics[width=0.8\linewidth]{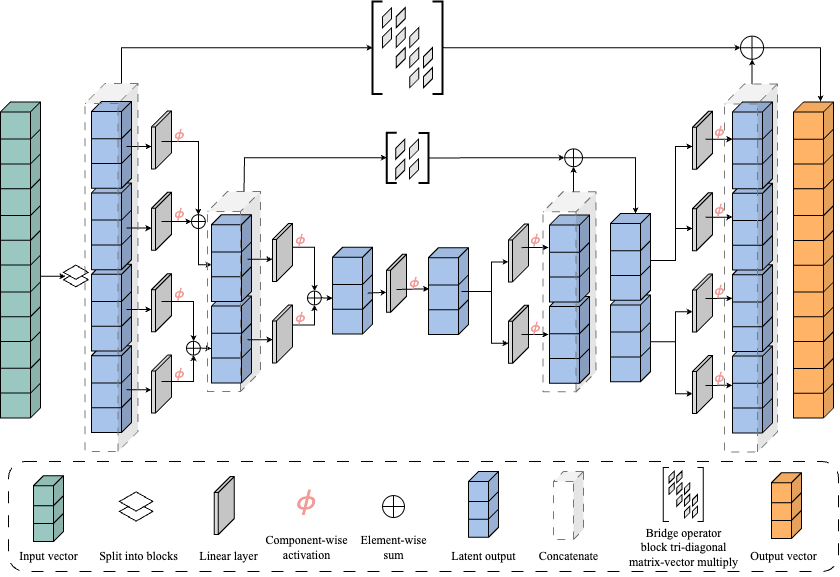}
    \caption{A schematic depiction of a one-dimensional GFMM-block. If the activation function $\phi$ is set to the identity map, the GFMM-block reduces to a linear block.}
    \label{fig:fmm_block}
\end{figure}

Discretization transforms the linear PDE into a linear system $Au = c$, with $A$ as the discretized forward operator. In the setting of PDE operators, it is well-known that $A$ often exhibits low-rank structures. Analytic results, such as in \citet{chandrasekaran2010numerical}, provide additional theoretical backing to these observations. Starting with foundational techniques such as FMM~\citep{greengard1987fast} and hierarchical matrices\citep{borm2003introduction}, various algebraic representations have been proposed to capture the low-rank structures in $A$ so that efficient linear algebra operations can be performed with them~\citep{chandrasekaran2003fast, xia2010fast}. While most representations have fast matrix-vector product algorithms, they do not always have fast solvers which are essential as preconditioners in iterative methods. In general, finding efficient low-rank representations for inverse operators of nonlinear PDEs using classical techniques is an open problem. Deep learning techniques provide an alternative framework, and rank-structured representations have been proposed in neural networks to learn inverse operators~\citep{kovachki2023neural,fan2019multiscale, Li2020MultipoleGNA, sushnikova2022fmm}.
\vspace{-0.2cm}
\paragraph{Linear GFMM}
We follow along similar lines and suggest a generalized FMM feed-forward architecture. Figure \ref{fig:fmm_block} illustrates a one-dimensional linear GFMM-block if $\phi$ is set to the identity. This block is an extension of the signal flow graph of the FMM matrix-vector product in Figure \ref{fig:fmm_sfg}, but with the parameters of the FMM signal flow graph replaced with learnable weights. Furthermore, general banded matrices are included as additional ``bridge'' operators.  As depicted in Figure \ref{fig:fmm_block}, the GFMM block contains three types of layers: down-sampling/encoder layers, up-sampling/decoder layers, and bridge layers. Similar to the original FMM method, the number of up-sampling(down-sampling) layers $L$ is a hyper-parameter for the network, and is typically chosen based on the physics of the underlying PDE and also memory constraints of the FMM solver. It defines the blocking size of the input vector and, consequently, the number of learnable parameters in the network. In its simplest form, all the layers are composed of $P \times P$ matrices, which apply linear transformations to their input vectors of size $P$, where $P=\frac{D}{2^L}$ and $D$ is the size of the input vector $c$.   
The full forward pass of GFMM-block is described in Algorithm \ref{alg:fmm_block}, where \texttt{BasisTransform($\mathbf{F}, y$)} corresponds to the basis transformation within the feed-forward network. For a single-channel GFMM-block where $y$ is a vector and $\mathbf{F}$ is a $P \times P$ linear layer, this operation corresponds to a matrix-vector multiply. We extend this to support multi-channel layers and inputs in Section \ref{sec:uno} and Algorithm \ref{alg:multi_domain}. A careful count reveals a total number of $(10\times 2^L -2L-9)P^2$ learnable parameters in the linear one-dimensional GFMM-block. To put this into contrast, a fully dense linear layer would involve $D^2 = 2^{2L}P^2$ parameters. The one-dimensional GFMM-block can be generalized to a two-dimensional one by applying analogous modifications to the 2D FMM signal flow graph which, in essence, is a tensor product of 1D signal flow graph (see Appendix \ref{sec:2d}).

\begin{algorithm}
\small
\caption{GFMM-block forward pass}\label{alg:fmm_block}
\begin{algorithmic}[1]
\begin{multicols}{2} 

\Input $c\in \mathbb{R}^D$
\Output $u\in \mathbb{R}^D$
\Parameters
    \State $L$: No. of up-sampling/down-sampling layers
    \State $M = 2^L$: Number of blocks
    \State $P = \frac{D}{M}$: Block size
    \State $\mathbf{E}^l_i, \mathbf{D}^l_i \in \mathbb{R}^{P \times P}$ for $l = 1,\ldots,L$
    \Statex and $i = 1,\ldots,\frac{M}{2^l}$: 
    \Statex Encoder and decoder layer parameters
    \State $\mathbf{B}^l \in \mathbb{R}^{\frac{M}{2^l}P \times \frac{M}{2^l}P}$ for $l = 0,\ldots,L$: 
    \Statex Bridge parameters
\columnbreak 
\scriptsize{
\Procedure{GFMM-Block}{$c$}
\State $h^0 \gets [c_1, c_2, \ldots, c_M]$
\For{$l=1$ to $L$}
    \State \hspace{-10pt} $h^l_i \gets $ \Call{BasisTransform}{}$\left(\mathbf{E}^l_{2i-1}, h^{l-1}_{2i-1}\right) + $
    \Statex \hspace{-10pt} \hspace{1cm} \Call{BasisTransform}{}$\left(\mathbf{E}^l_{2i}, h^{l-1}_{2i}\right)$ for $i=1$ to $\frac{M}{2^l}$
    \State \hspace{-10pt} $h^l \gets [h^l_1, h^l_2,\ldots,h^l_{\frac{M}{2^l}}]$
\EndFor
\State $z^L \gets \mathbf{B}^Lh^L$
\For{$l=L-1$ to $0$}
    \State \hspace{-10pt} $z^l_i \gets $ \Call{BasisTransform}{}$\left(\mathbf{D}^l_i,  z^{l+1}_{\floor{\frac{i+1}{2}}}\right)$ for $i=1$ to $\frac{M}{2^l}$
    \State \hspace{-10pt} $z^l \gets [z^l_1, z^l_2,\ldots,z^l_{\frac{M}{2^l}}]$
    \State \hspace{-10pt} $z^l \gets z^l + \mathbf{B}^lh^l$
\EndFor
\State $u \gets z^0$
\EndProcedure}
\end{multicols} 
\end{algorithmic}
\end{algorithm}

\vspace{-0.5cm}
\paragraph{Nonlinear GFMM}
While linear GFMM-blocks are suitable architectures to represent inverse operators of linear PDEs, they have to be extended with nonlinear activations to model nonlinear PDEs. Nonlinear activations are also required to map coefficients $a,f$ of the PDE to the solution since this relationship is nonlinear by nature. We use the nonlinear rational function $\phi (x) = \frac{x}{1+|x|}$ as the activation function, which is motivated by the observation that the solution to a linear partial differential equation is a rational function of its coefficients. Later in our experiments (see section \ref{sec:relu_vs_rational}), we show that this type of rational activation function seems to perform better than ReLUs.

\subsection{Uni-modal Neural Operator (UNO) with GFMM-blocks} \label{sec:uno}

The primary motivation for using the GFMM-block architecture is the efficient representation of inverse operators of rank-structured linear systems. But learning the representation of a generic PDE using a single GFMM-block might be too restrictive. Experimentally, we observed that sequentially stacking multiple GFMM blocks leads to faster convergence of the training error. This is not surprising given the benefits of depth in neural networks~\citep{telgarsky2016benefits} and the increase in the number of learnable parameters. Subsequently, we define the uni-modal neural operator (UNO) as a stack of GFMM-blocks with the number of blocks being a hyper-parameter that can be tuned. Moreover, to enable learning the solution operator for more than one input parameter, we extend the GFMM-blocks in the UNO architecture to support multi-channel inputs and outputs. We do this by replacing each $P\times P$ linear layer $\mathbf{F}$ in the encoder and decoder layers by a 4-dimensional tensor of size $C_{out} \times C_{in} \times P \times P$, where $C_{out}$ is the number of output channels desired and $C_{in}$ is the number of channels in the input. The linear transformation is applied such that each channel in the output vector $z$ is obtained by a matrix-vector multiplication of input tensor $y$ with the $P \times P$ weight matrix of the corresponding channel and them summed over all the input channels i.e. $z(i) = \sum_j \mathbf{F}(i,j,:,:) y(j)$.
\vspace{-0.2cm}
\subsection{Multi-modal Fusion}\label{sec: multi_fusion}
\begin{wrapfigure}{r}{0.42\textwidth}
    \centering
    \includegraphics[width=0.4\textwidth]{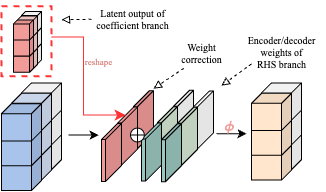}
    \caption{Multi-modal fusion: Basis transformation in one layer of the RHS-branch of MNO. Complete multi-modal fusion across the entire MNO is shown in Figure \ref{fig:mlnet} in the Appendix.}
    \label{fig:multi_fusion}
\end{wrapfigure}

Similar to FNO and DeepONet, UNO can approximate the solution operator of PDEs that map either the coefficients or the RHS to the solution, while keeping other parameters fixed. To extend UNO to a complete MNO network involves the construction in Figure \ref{fig:mno_overview} with branches of stacked GFMM-blocks corresponding, respectively, to the coefficient and RHS-branch. The RHS-branch is analogous to a UNO that directly maps the RHS to the solution, but in order to learn the simultaneous map from both coefficients and RHS to the solution, we ``fuse'' the weights of the RHS-branch using the latent outputs produced by GFMM-blocks of the coefficient branch.  We describe this fusing operation as \emph{multi-modal fusion} as illustrated in Figure \ref{fig:multi_fusion}. The rationale behind this operation can be explained as follows. Consider a simple linear PDE described by
\begin{eqnarray*}
   a(x)u'(x) + u(x) = c(x).
\end{eqnarray*}
For the discrete version of this problem the linear operator (including BC) that maps $c$ to $u$ is a rational function of the sampled values of $a$. Moreover this rational dependence is also true for the FMM representation of the inverse, though that fact requires a careful perusal of the literature, but is known to experts. Since it is well-known that deep networks can generate rational functions effectively, it makes sense to use a separate DNN to generate the FMM coefficients for the inverse from the samples of $a$. Furthermore, by doing a perturbation analysis, one can show that localized changes to $a$ will primarily affect the solution $u$ in the same location. Motivated by this reasoning it also makes sense to posit that the DNN that maps $a$ to the FMM weights, itself should have an FMM-like architecture.
\vspace{2cm}
\section{Experiments} \label{sec:experiments}
\begin{wraptable}{r}{0.65\textwidth}
\scriptsize
\centering
\vspace{-0.5cm}
\caption{Summary of the experiments setup.}
\label{tab:equation_comparison_with_headings}
{\setlength{\tabcolsep}{2pt}
{\renewcommand{\arraystretch}{1.3}
\begin{tabular}{lccc}
\hline
\textbf{} & \textbf{Poisson's} & \textbf{Darcy flow} & \textbf{Generic first order} \\
\hline
Equations & $\begin{gathered}
            A u = c \\
           A : \text{1D Laplacian}
            \end{gathered}$
          & $\begin{gathered}
            -\nabla. (a \nabla u) = c \\
            u(0) = u_0; u(1) = u_1
            \end{gathered}$
          & $\begin{gathered}
            a(x)u' + b(x)|u| = c(x) \\
            \int f(x)u(x)dx = g
            \end{gathered}$ \\
\hline
Linearity & linear & linear & nonlinear \\
\hline
Coefficient & constant coeff & single variable coeff & multiple variable coeffs \\
\hline
BC & fixed BC & fixed BC & parametric integral BC \\
\hline
Map & $c \mapsto u$ & $a,c \mapsto u$ & $a,b,c,f,g \mapsto u$ \\
\hline
Model & UNO & MNO & MNO \\
\bottomrule 
\end{tabular}}}
\vspace{-0.8cm}
\end{wraptable}

In this section, we train and evaluate the performance of our UNO and MNO networks on both linear and nonlinear BVPs, emphasizing their ability to learn solution operators across different configurations, as shown in Table \ref{tab:equation_comparison_with_headings}. We also focus on the out-of-distribution performance of the models trained using synthetic data.

We begin by showing that our linear UNO network can approximate the inverse operator of discrete 1D Poisson's equation. We compare this with FNO and DeepONet in learning the map between $c(x)$ and $u(x)$.
Next we show our MNO network learning the joint map $a,c \mapsto u$ for the case of 1D Darcy flow with fixed boundary conditions.
Finally, we present the experiments for the case of the full map between all the parameters of the PDE and the boundary, for multiple RHS for a generic nonlinear BVP.
\vspace{-0.2cm}
\subsection{Experimental setup} \label{sec:exp_setup}
\paragraph{Training data}
We adopt the synthetic data approach from \citet{hasani2024generating}, constructing the synthetic solution $u$ using the Chebyshev basis: $u(x) = \sum^{N_{\alpha}}_{k=1} \alpha_k T_n(x)$, where $T_n$ are the Chebyshev polynomials of the first kind ($T_n(\cos x) = \cos(nx)$), and $\alpha_k$ is uniformly sampled from $[-1,1]$. The specific choice of PDE coefficients ($a,f$) for each problem are described in their respective sections. We use finite difference discretization of the operators $\mathcal{D}_1$ and $\mathcal{D}_2$ in (\ref{eqn:gen_pde}) to generate $c$ and $g$ for a particular synthetic $u$. We resample $\alpha_k$ at every iteration, which serves as a regularization strategy to mitigate overfitting. We refer to this scheme as \emph{solution sampling}.

\paragraph{Validation data}
We employ two types of validation data to evaluate the out-of-distribution generalization of our models. In the first type, we sample coefficient parameters from distributions that are different from the training distribution, while keeping $u$ drawn from the same distribution, to test generalization across coefficient function spaces for $a$. In the second, we instead sample the RHS by constructing $c(x)= \sum^{N_{\varphi}}_{k=1} \varphi_k T_n(x)$, with $\varphi_k$ sampled from $\mathcal{U}[-1,1]$ and evaluated on the same grid. We refer to this scheme as \emph{RHS sampling} and evaluate the models using the residual error (\ref{eqn:res_err}). While most neural operator methods adopt RHS sampling with classical solvers to obtain ground-truth $u$ for training and validation, such solvers are not easily available for complex nonlinear PDEs, and can prove very costly to operate when they are. Instead, we evaluate generalization on RHS-sampled data, while using only solution-sampled data for training.

\paragraph{Training method}
We use a grid of $256$ points and choose the Chebyshev basis set as $T_1$ to $T_{16}$. We use single precision and the mean squared error of the predicted solution $\hat{u}$ as the loss function and train using the Adam~\citep{kingma2014adam} optimizer with an initial learning rate of 1e-03. We used two NVIDIA TITAN RTX GPUs for all the experiments.

\paragraph{Error metrics}
For the linear constant coefficient case, where solving the discretized PDE is equivalent to solving the linear problem $Au = c$ where $A$ is a $D \times D$ matrix, we define the normalized backward error $ \epsilon_{\text{be}}$ as in (\ref{eqn:res_err}),
where $\hat{u}$ is the model's predicted solution, and the norm is the standard 2-norm.
For the variable coefficient and nonlinear case, we measure the residual error, $ \epsilon_{\text{res}}$, of the forward operator $\mathcal{D}$ applied on the model's predicted solution $\hat{u}$, and the relative solution error, $\epsilon_{\text{rel}}$:
\begin{eqnarray}
    \epsilon_{\text{be}} = \frac{\|A\hat{u}-c\|}{\|A\| \|\hat{u}\| + \|c\|}; \qquad
    \epsilon_{\text{res}} = \| \mathcal{D}(x,\hat{u}) - c(x) \|; \qquad
    \epsilon_{\text{rel}} = \frac{\| \hat{u} - u \|}{\| u \|}. \label{eqn:res_err}
\end{eqnarray}

\subsection{Linear PDEs}

\subsubsection{Uni-modal: Discrete 1D Poisson's} \label{sec:p1d}
We train UNO network to approximate the inverse operator of a linear set of equations given by $Au = c$, where $A$ is the discrete 1D Laplacian operator.
Our UNO model is composed of two linear GFMM-blocks with parameters $P=16, L=4$ sequentially stacked. To compare the performance of UNO, we also train an FNO with \texttt{width=32}, \texttt{modes=16} and a DeepONet with 3 branch layers and 3 trunk layers, each with \texttt{width=128}. We train all the models for 120k iterations starting with a learning rate of 1e-03 and dropping to 1e-04 after 20k iterations on \emph{only} the Chebyshev basis polynomials, i.e., $u_i = \alpha_k T_k$. 

\begin{figure}
    \centering
    \includegraphics[width=\linewidth]{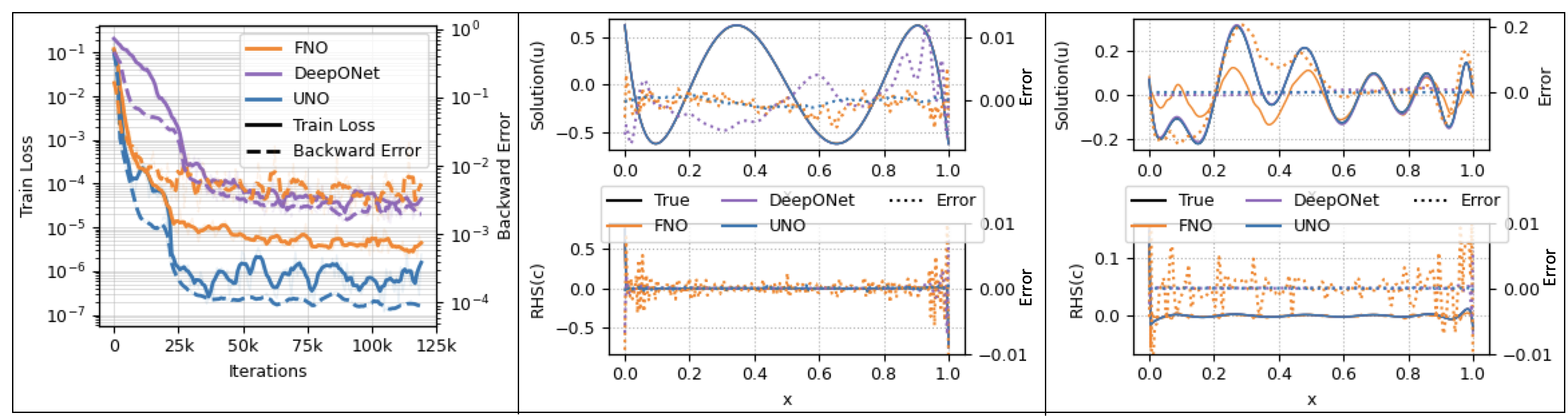}
    \caption{(Left) Training loss and backward error during training for 1D Poisson models. Predicted vs true solutions and corresponding RHS on randomly scaled (middle) and random linear combinations (right) of Chebyshev polynomials representing in-training and out-of-training distribution. The dotted lines represent the error in predictions for each model.}
    \label{fig:p1d_losses}
\end{figure}

\begin{table}
    \small
      \centering
      {\setlength{\tabcolsep}{3pt}
      {
    \caption{Relative error and residual error of FNO, DeepONet and UNO for 1D Poisson equation. I) For ``Solution sampling'' columns, $u$ is i) a random $T_k$ (Chebyshev basis), scaled by random $\alpha_k$, which is same as training data, and ii) random linear combinations of $T_k$ which is out-of-training distribution (OOD). II) For ``RHS sampling'' columns, $c$ is i) a random $T_k$ (Chebyshev basis), scaled by random $\alpha_k$ and ii) random linear combinations of $T_k$, but in this case both constitute out-of-training distribution.}
    \label{tab:p1d_results}
    \begin{tabular}{lccccccccc}
        \toprule
        \multirow{3}{*}{Model} & \multirow{3}{*}{\shortstack{Number \\of params}} & \multicolumn{4}{c}{Solution sampling} & \multicolumn{4}{c}{RHS sampling} \\
        \cmidrule(lr){3-6} \cmidrule(lr){7-10}
        & & \multicolumn{2}{c}{$u = \alpha_k T_k$(Train)} & \multicolumn{2}{c}{$u = \sum_k\alpha_k T_k$(OOD)} & \multicolumn{2}{c}{$c = \alpha_k T_k$(OOD)} & \multicolumn{2}{c}{$c = \sum_k \alpha_k T_k$(OOD)} \\
        \cmidrule(lr){3-4} \cmidrule(lr){5-6} \cmidrule(lr){7-8} \cmidrule(lr){9-10}
        & & $\epsilon_{\text{rel}}$ & $\epsilon_{\text{be}}$ & $\epsilon_{\text{rel}}$ & $\epsilon_{\text{be}}$ & $\epsilon_{\text{rel}}$ & $\epsilon_{\text{be}}$ & $\epsilon_{\text{rel}}$ & $\epsilon_{\text{be}}$ \\
        \midrule
        FNO & 139,713 & 8.24e-03 & 1.72e-3 & 4.54e-01 & 3.29e-2 & 7.56e-01 & 1.29e-2 & 9.63e-01 & 2.15e-2 \\
        DeepONet & 132,224 & 2.05e-03 & 1.60e-3 & 1.09e-03 & 2.40e-3 & 2.63e-01 & 1.98e-2 & 7.31e-01 & 1.45e-2 \\
        UNO & \textbf{73,216} & \textbf{5.32e-06} & \textbf{1.12e-4} & \textbf{5.72e-06} & \textbf{1.14e-4} & \textbf{8.18e-02} & \textbf{2.08e-3} & \textbf{5.76e-01} & \textbf{4.78e-3} \\
        \bottomrule
    \end{tabular}}}
\end{table}

Figure \ref{fig:p1d_losses} (left) illustrates that while all models achieve a training loss below 1e-4, UNO achieves a backward error $\epsilon_{\text{be}}$ approximately two orders of magnitude lower than FNO and DeepONet, despite having roughly half their number of learnable parameters. As seen from Table \ref{tab:p1d_results}, though all models accurately predict when $u$ is a randomly scaled Chebyshev polynomial similar to the training set (see Figure \ref{fig:p1d_losses} (middle)), UNO demonstrates better generalization compared to FNO and DeepONet when $u$ is sampled from out-of-training distribution (see Figure \ref{fig:p1d_losses} (right)). But when tested on data generated via RHS sampling (where $c$ is sampled and $u$ is derived), all models struggled to generalize, although UNO shows a slight advantage, particularly in its lower backward error (also see Figure \ref{fig:p1d_mixcheb1}).

\subsubsection{Multi-modal: 1D Darcy Flow} \label{sec:darcy}
For the multi-modal case, we consider the steady state Darcy flow in 1D with variable coefficient $a(x)$ and fixed Dirichlet boundary conditions:
\begin{equation*}
    -\nabla (a(x) \nabla u(x)) = c(x),\quad  0 < x < 1, \quad u(0) = u_0, \quad u(1) = u_1. 
\end{equation*}
Our goal is to learn the solution operator $a,c \mapsto u$.
We use our multi-modal network MNO with one channel inputs for both coefficient and RHS-branches, while the boundary conditions are fixed. We choose $L=4$ and $P=16$ as the network's parameters. $a$ is sampled from a mixture distribution of quadratic parametric form and log-normal distributions:
\begin{subequations} 
\begin{align}
    a(x) &= 0.5a_1(x) + 0.5 a_2(x); \label{eqn:darcy_a_distr_nc} \\
    a_1(x) &= 1 + \theta_1 x^2; \quad \text{where }  \theta_1 \sim \mathcal{U}(0,1) \label{eqn:darcy_a_case1_nc}, \\ 
    a_2(x) &= 0.1 + e^{{\theta_2}}; \quad \text{where } \theta_2 \sim \mathcal{N}(0,k(x,x')), \; k(x,x') = e^{ -\frac{|x-x'|^2}{2(0.1D)^2}}. \label{eqn:darcy_a_case2_nc}  
\end{align}
\end{subequations}
In subsurface flow modeling, e.g., in groundwater flow or oil reservoir simulations governed by steady-state Darcy's law, the permeability field $a(x)$ typically exhibits high heterogeneity. Geological layers formed by sediment deposition have spatial correlations captured by log-normal gaussian processes whereas certain smooth layers are approximated by polynomial spatial variability and learning to map such heterogeneous distributions is useful in practice.

\begin{figure}
    \centering
    \includegraphics[width=\linewidth]{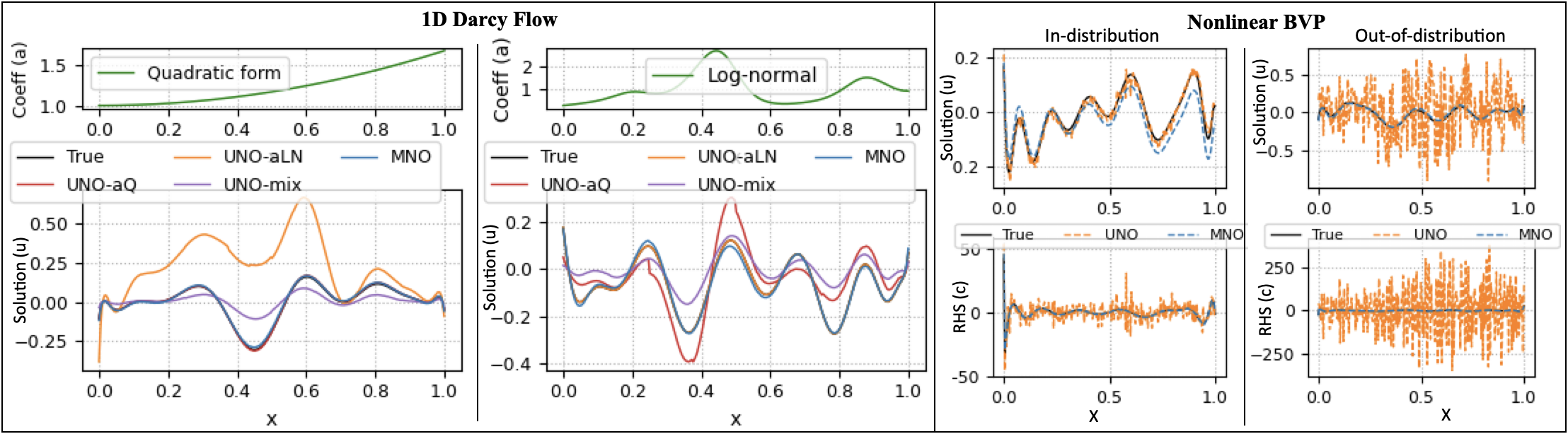}
    \caption{(Left) Test samples for 1D Darcy flow where the coefficient $a$ is drawn from (\ref{eqn:darcy_a_case1_nc}) and (\ref{eqn:darcy_a_case2_nc}) shows that MNO models can learn the simultaneous map $a,c \mapsto u$ while the UNO models cannot. (Right) MNO learns the solution map from multiple parameters but UNO struggles for both in-distribution and out-of-distribution coefficients.}
    \label{fig:darcy1d}
\end{figure}

We evaluate MNO against three UNO networks. UNO-aQ and UNO-aLN were trained with a fixed coefficient $a$ sampled from distinct distributions, (\ref{eqn:darcy_a_case1_nc}) \& (\ref{eqn:darcy_a_case2_nc}) respectively, while UNO-mix encountered a new $a$ (\ref{eqn:darcy_a_distr_nc}) in each training iteration, similar to MNO. Unlike MNO, which receives both coefficient $a$ and the RHS $c$, UNOs only accept $c$ as input. All models achieved a training MSE below 1e-3 over 50k iterations with a mini-batch size of 1000.
Figure \ref{fig:darcy1d}(left) demonstrates that while UNO-aQ and UNO-aLN only succeeded for their specific training $a$ and UNO-mix struggled with novel $a$ values despite its varied training exposure, MNO generalized effectively to unseen coefficients. Furthermore, Table \ref{tab:darcy1d}, with results averaged over 1000 solution sampled test examples, shows that uni-modal networks failed to adapt to changes in $a$, whereas MNO robustly mapped both $a$ and $c$ simultaneously to the solution.

\begin{table}
\small
    \centering
    \caption{Relative error and residual error of UNO and MNO models trained on 1D Darcy flow with coefficients $a$ sampled purely from either (\ref{eqn:darcy_a_case1_nc}) or (\ref{eqn:darcy_a_case2_nc}). }
    \label{tab:darcy1d}
    \begin{tabular}{c|cc|cc}
        \hline
        \multirow{2}{*}{Model} & \multicolumn{2}{c|}{$a \in$ (\ref{eqn:darcy_a_case1_nc})(Quadratic)} & \multicolumn{2}{c}{$a \in$ (\ref{eqn:darcy_a_case2_nc})(Log-normal)} \\
        \cline{2-5}
         & $\epsilon_{\text{rel}}$ & $\epsilon_{\text{res}}$ & $\epsilon_{\text{rel}}$ & $\epsilon_{\text{res}}$ \\
        \hline
        UNO-aQ & (2.71$\pm$0.04)e-02 & (6.86$\pm$0.07)e-04 & (4.07$\pm$0.02)e+00 & (1.74$\pm$0.55)e-02 \\
        UNO-aLN & (1.22$\pm$0.002)e+01 & (3.76$\pm$0.03)e-03 & (1.95$\pm$0.002)e+01 & (1.73$\pm$0.09)e-02 \\
        UNO-mix & (4.94$\pm$0.01)e-01 & (1.96$\pm$0.04)e-03 & (5.86$\pm$0.05)e-01 & (5.39$\pm$0.12)e-03 \\
        MNO & \textbf{(4.42$\pm$0.03)e-03} & \textbf{(3.77$\pm$0.06)e-04} & \textbf{(3.37$\pm$0.07)e-02} & \textbf{(1.53$\pm$0.01)e-03} \\
        \hline
    \end{tabular}
\end{table}

\subsection{Nonlinear ODEs}
In the section we use the following nonlinear first order BVP with a parametric form:
\begin{subequations}  \label{eq:integralequation}
\begin{align}
    a(x)u'(x) + b(x)|u(x)| = c(x) ; \quad  & 0<x<1, \label{eq:full_mm_int} \\
    \int f(x)u(x)dx = g; \quad  0 \le x \le 1 \label{eq:full_mm_bnd}
\end{align}
\end{subequations}
where the boundary condition (\ref{eq:full_mm_bnd}) is also given in a general parametric form. The absolute value of the solution term and an additional variable coefficient $b(x)$ are chosen only to test model's ability to handle nonlinearity and multiple coefficient parameters, which are sampled from following distributions:
\begin{eqnarray}
    \mathcal{P}_{\text{tr}} : \begin{cases}
        a =  1 + \theta x^2; \\
        b = \sum_{i=1}^{4} \phi_{i} T_i(x) ; \\
        f =\sum_{i=1}^{4} \eta_i T_i(x) ; \text{where } \theta, \phi_i, \eta_i \sim \mathcal{U}[-1,1] .
    \end{cases}
\end{eqnarray}

\subsubsection{Uni-modal}
We train a UNO network with nonlinear activation where all the parameters $a, b, c, f, g$ are stacked as multiple channels of the input to the UNO for 20k iterations. The GFMM-blocks of UNO are initialized with $P=16$ and $L=4$. We evaluate the network using coefficients sampled from both in-distribution and out-of-distribution on batches of 1000 test samples. Table \ref{tab:full_mm} shows the relative error, interior residual error and boundary residual error on different combinations of the sampled test data distributions. UNO achieves low relative error in the predicted solution but higher residual errors with in-distribution test data, suggesting overfitting on the solution without actually learning the joint map from the parameters. Moreover, the UNO model struggles further with out-of-distribution data, failing to attain even low relative error. Examples of UNO model's predictions in both cases are shown in Figure \ref{fig:darcy1d}(right), where it becomes evident that UNO is unable to smoothly approximate the join map from multiple parameters to the solution. Next we discuss our multi-modal network's performance on the same setup.

\begin{table}
\scriptsize
  \centering
  \caption{Relative error, interior residual error and boundary residual error of UNO and MNO models for nonlinear BVP in (\ref{eq:integralequation}). The errors are measured on 1000 test samples by sampling interior coefficients ($a,b$) and boundary coefficient ($f$) either from $\mathcal{P_{\text{tr}}}$ or from a unit normal (out-of-training) distribution.}
  \label{tab:full_mm}
  {\setlength{\tabcolsep}{3pt}
  {\renewcommand{\arraystretch}{1.3}
  \begin{tabular}{c|c|ccc|ccc}
    \hline
    \multicolumn{2}{c|}{\multirow{2}{*}{}} & \multicolumn{3}{c|}{$a,b \sim \mathcal{P_{\text{tr}}}$} & \multicolumn{3}{c}{$a,b \nsim \mathcal{P_{\text{tr}}}$} \\
    \multicolumn{2}{c|}{} & $\epsilon_{\text{rel}}$ & $\epsilon_{\text{res}}^{\text{int}}$ & $\epsilon_{\text{res}}^{\text{bnd}}$ & $\epsilon_{\text{rel}}$ & $\epsilon_{\text{res}}^{\text{int}}$ & $\epsilon_{\text{res}}^{\text{bnd}}$ \\
    \hline
    \multirow{2}{*}{\centering $f \sim \mathcal{P_{\text{tr}}}$} & 
    UNO & (4.1$\pm$0.07)e-02 & (4.8$\pm$0.04)e+00 & (1.4$\pm$0.05)e-02 & (1.7$\pm$0.02)e+01 & (1.2$\pm$0.08)e+02 & (5.6$\pm$0.03)e-02 \\
    & MNO & (7.0$\pm$0.02)e-02 & (3.8$\pm$0.06)e-01 & (8.4$\pm$0.09)e-03 & (8.0$\pm$0.05)e-02 & (4.4$\pm$0.01)e-01 & (6.0$\pm$0.07)e-03 \\
    \hline
    \multirow{2}{*}{\centering $f \nsim \mathcal{P_{\text{tr}}}$} & 
    UNO & (1.1$\pm$0.03)e+01 & (1.2$\pm$0.09)e+02 & (2.5$\pm$0.04)e-02 & (2.8$\pm$0.06)e+01 & (1.9$\pm$0.001)e+02 & (7.3$\pm$0.01)e-02 \\
    & MNO & (1.1$\pm$0.01)e-01 & (4.2$\pm$0.37)e-01 & (1.6$\pm$0.30)e-03 & (1.3$\pm$0.09)e-01 & (5.5$\pm$0.14)e-01 & (1.2$\pm$0.42)e-03 \\
    \hline
  \end{tabular}}}
\end{table}

\vspace{-0.2cm}
\subsubsection{Multi-modal} \label{sec:full_mm}
We use the same nonlinear BVP in (\ref{eq:integralequation}) to train our MNO that learns the solution map not only from the parameters of the interior equations ($a, b, c$) but also boundary parametric constraints ($f, g$).

MNO is initialized with two nonlinear GFMM-blocks in the coefficient branch and one nonlinear and one linear GFMM-blocks in the RHS-branch with each block parameterized using $P=16$ and $L=4$. 
Figure \ref{fig:darcy1d} (right) shows that MNO's predicted solution is smoother and generalizes better to out-of-distribution coefficients. Lower residual errors across all distributions compared to UNO, as seen from Table \ref{tab:full_mm}, indicates that MNO is able to learn the solution operator mapping not only the coefficients and RHS but also the parameterized boundary conditions with better generalization.

\vspace{-0.2cm}
\subsection{Out-of-distribution Generalization} \label{sec:soln_rhs_emp}
Table \ref{tab:soln_vs_rhs} shows the mean residual errors for each experiment when the models are tested on 1000 RHS sampled examples, which are out-of-training distribution. We note the large increase in error on RHS. To fix this, researchers have proposed that the output of these networks to be refined by classical iterative solvers~\citep{zhang2024blending}.

\begin{table}
\small
    \centering
    \caption{Residual errors computed on test data sampled using solution sampling (SS) and RHS sampling for 1D Poisson's, 1D Darcy flow and BVP of (\ref{eq:integralequation}) for different models.}
    \label{tab:soln_vs_rhs}
    \begin{subtable}[t]{0.31\textwidth}
        \centering
        \caption{Poisson's}
        {\setlength{\tabcolsep}{2pt}
          \begin{tabular}{lcc}
            \toprule
            {} & SS & RHS \\
            \midrule
            FNO & 3.50e-03 & \textbf{1.48e-01} \\
            DeepONet & 1.90e-04 & 3.68e-01 \\
            UNO & \textbf{2.83e-05} & 4.96e-01 \\
          \end{tabular}}
          \label{tab:p1d_soln_rhs}
    \end{subtable}
    \begin{subtable}[t]{0.31\textwidth}
        \centering
        \caption{Darcy flow}
        {\setlength{\tabcolsep}{2pt}
        \begin{tabular}{lcc}
            \toprule
            {} & SS & RHS \\
            \midrule
            UNO-aQ & 4.26e-04 & 2.64e-01 \\
            UNO-aLN & 3.99e-03 & 1.88e-01 \\
            UNO-mix & 1.98e-03 & 6.19e-01 \\
            MNO & \textbf{3.93e-04} & \textbf{9.22e-02} \\
      \end{tabular}}
      \label{tab:darcy_soln_rhs}
    \end{subtable}
    \begin{subtable}[t]{0.31\textwidth}
        \centering
        \caption{BVP}
        {\setlength{\tabcolsep}{2pt}
        \begin{tabular}{lcc}
            \toprule
            {} & SS & RHS \\
            \midrule
            UNO & 6.10e+00 & 1.12e+02 \\
            MNO & \textbf{4.12e-01} & \textbf{2.97e+00} \\
          \end{tabular}}
          \label{tab:full_mm_soln_rhs}
    \end{subtable}
\end{table}
\vspace{-0.2cm}
\section{Conclusions \& Future work} \label{sec:conclusion}
We presented the Multi-modal Neural Operator (MNO) for learning solution operators of parameterized PDEs under simultaneous variations in coefficients, source terms, and boundary conditions. While our current demonstrations are primarily 1D (with preliminary 2D results in Appendix \ref{sec:2d}), future work will focus on extending MNO to higher dimensions, time-evolving PDEs, and diverse domain geometries. The demonstrated multi-modal capabilities of MNO also suggest its potential in applications like reducing the computational cost of neural-assisted preconditioners for classical iterative solvers. We believe our preliminary results will motivate the community to further explore and develop multi-modal architectures, significantly broadening the scope and utility of neural operators in scientific computing and engineering applications.



{
\small
\bibliography{refs}

\begin{thebibliography}{38}
\providecommand{\natexlab}[1]{#1}
\providecommand{\url}[1]{\texttt{#1}}
\expandafter\ifx\csname urlstyle\endcsname\relax
  \providecommand{\doi}[1]{doi: #1}\else
  \providecommand{\doi}{doi: \begingroup \urlstyle{rm}\Url}\fi

\bibitem[Aldirany et~al.(2024)Aldirany, Cottereau, Laforest, and Prudhomme]{aldirany2024multi}
Z.~Aldirany, R.~Cottereau, M.~Laforest, and S.~Prudhomme.
\newblock Multi-level neural networks for accurate solutions of boundary-value problems.
\newblock \emph{Computer Methods in Applied Mechanics and Engineering}, 419:\penalty0 116666, 2024.

\bibitem[B{\"o}rm et~al.(2003)B{\"o}rm, Grasedyck, and Hackbusch]{borm2003introduction}
S.~B{\"o}rm, L.~Grasedyck, and W.~Hackbusch.
\newblock Introduction to hierarchical matrices with applications.
\newblock \emph{Engineering analysis with boundary elements}, 27\penalty0 (5):\penalty0 405--422, 2003.

\bibitem[Cai et~al.(2021)Cai, Wang, Lu, Zaki, and Karniadakis]{cai2021deepm}
S.~Cai, Z.~Wang, L.~Lu, T.~A. Zaki, and G.~E. Karniadakis.
\newblock Deepm\&mnet: Inferring the electroconvection multiphysics fields based on operator approximation by neural networks.
\newblock \emph{Journal of Computational Physics}, 436:\penalty0 110296, 2021.

\bibitem[Chandrasekaran et~al.(2003)Chandrasekaran, Dewilde, Gu, Pals, Sun, van~der Veen, and White]{chandrasekaran2003fast}
S.~Chandrasekaran, P.~Dewilde, M.~Gu, T.~Pals, X.~Sun, A.~van~der Veen, and D.~White.
\newblock Fast stable solvers for sequentially semi-separable linear systems of equations and least squares problems.
\newblock \emph{SIAM Journal on Matrix Analysis and Applications}, 2003.

\bibitem[Chandrasekaran et~al.(2010)Chandrasekaran, Dewilde, Gu, and Somasunderam]{chandrasekaran2010numerical}
S.~Chandrasekaran, P.~Dewilde, M.~Gu, and N.~Somasunderam.
\newblock On the numerical rank of the off-diagonal blocks of schur complements of discretized elliptic pdes.
\newblock \emph{SIAM Journal on Matrix Analysis and Applications}, 31\penalty0 (5):\penalty0 2261--2290, 2010.

\bibitem[Chen et~al.(2024{\natexlab{a}})Chen, Liu, Lin, Chen, and Shi]{nsno2024}
F.~Chen, Z.~Liu, G.~Lin, J.~Chen, and Z.~Shi.
\newblock Nsno: Neumann series neural operator for solving helmholtz equations in inhomogeneous medium.
\newblock \emph{Journal of Systems Science and Complexity}, 37\penalty0 (2):\penalty0 413--440, 2024{\natexlab{a}}.

\bibitem[Chen et~al.(2024{\natexlab{b}})Chen, Song, Ren, Subramanian, Morozov, and Mahoney]{chen2024data}
W.~Chen, J.~Song, P.~Ren, S.~Subramanian, D.~Morozov, and M.~W. Mahoney.
\newblock Data-efficient operator learning via unsupervised pretraining and in-context learning.
\newblock \emph{Advances in Neural Information Processing Systems}, 37:\penalty0 6213--6245, 2024{\natexlab{b}}.

\bibitem[Fan et~al.(2019)Fan, Feliu-Faba, Lin, Ying, and Zepeda-N{\'u}nez]{fan2019multiscale}
Y.~Fan, J.~Feliu-Faba, L.~Lin, L.~Ying, and L.~Zepeda-N{\'u}nez.
\newblock A multiscale neural network based on hierarchical nested bases.
\newblock \emph{Research in the Mathematical Sciences}, 6\penalty0 (2):\penalty0 21, 2019.

\bibitem[G{\'o}mez-Castro(2024)]{gomez2024beginner}
D.~G{\'o}mez-Castro.
\newblock Beginner’s guide to aggregation-diffusion equations.
\newblock \emph{SeMA Journal}, 81\penalty0 (4):\penalty0 531--587, 2024.

\bibitem[Greengard and Rokhlin(1987)]{greengard1987fast}
L.~Greengard and V.~Rokhlin.
\newblock A fast algorithm for particle simulations.
\newblock \emph{Journal of computational physics}, 73\penalty0 (2):\penalty0 325--348, 1987.

\bibitem[Gupta et~al.(2021)Gupta, Xiao, and Bogdan]{Gupta2021MultiwaveletbasedOLA}
G.~Gupta, X.~Xiao, and P.~Bogdan.
\newblock Multiwavelet-based operator learning for differential equations.
\newblock In \emph{Neural Information Processing Systems}, 2021.
\newblock URL \url{https://arxiv.org/pdf/2109.13459.pdf}.

\bibitem[Hao et~al.(2023)Hao, Wang, Su, Ying, Dong, Liu, Cheng, Song, and Zhu]{hao2023gnot}
Z.~Hao, Z.~Wang, H.~Su, C.~Ying, Y.~Dong, S.~Liu, Z.~Cheng, J.~Song, and J.~Zhu.
\newblock Gnot: A general neural operator transformer for operator learning.
\newblock In \emph{International Conference on Machine Learning}, pages 12556--12569. PMLR, 2023.

\bibitem[Hasani and Ward(2024)]{hasani2024generating}
E.~Hasani and R.~A. Ward.
\newblock Generating synthetic data for neural operators.
\newblock \emph{arXiv preprint arXiv:2401.02398}, 2024.

\bibitem[Helwig et~al.(2023)Helwig, Zhang, Fu, Kurtin, Wojtowytsch, and Ji]{Helwig2023GroupEFA}
J.~Helwig, X.~Zhang, C.~Fu, J.~Kurtin, S.~Wojtowytsch, and S.~Ji.
\newblock Group equivariant fourier neural operators for partial differential equations.
\newblock In \emph{International Conference on Machine Learning}, 2023.
\newblock URL \url{https://api.semanticscholar.org/CorpusId:259129662}.

\bibitem[Kingma(2014)]{kingma2014adam}
D.~P. Kingma.
\newblock Adam: A method for stochastic optimization.
\newblock \emph{arXiv preprint arXiv:1412.6980}, 2014.

\bibitem[Kovachki et~al.(2023)Kovachki, Li, Liu, Azizzadenesheli, Bhattacharya, Stuart, and Anandkumar]{kovachki2023neural}
N.~Kovachki, Z.~Li, B.~Liu, K.~Azizzadenesheli, K.~Bhattacharya, A.~Stuart, and A.~Anandkumar.
\newblock Neural operator: Learning maps between function spaces with applications to pdes.
\newblock \emph{Journal of Machine Learning Research}, 24\penalty0 (89):\penalty0 1--97, 2023.

\bibitem[Lerer et~al.(2024)Lerer, Ben-Yair, and Treister]{lerer2024multigrid}
B.~Lerer, I.~Ben-Yair, and E.~Treister.
\newblock Multigrid-augmented deep learning preconditioners for the helmholtz equation using compact implicit layers.
\newblock \emph{SIAM Journal on Scientific Computing}, 46\penalty0 (5):\penalty0 S123--S144, 2024.

\bibitem[Li et~al.(2021)Li, Kovachki, Azizzadenesheli, Liu, Bhattacharya, Stuart, and Anandkumar]{fno2021}
Z.~Li, N.~Kovachki, K.~Azizzadenesheli, B.~Liu, K.~Bhattacharya, A.~Stuart, and A.~Anandkumar.
\newblock Fourier neural operator for parametric partial differential equations.
\newblock In \emph{International Conference on Learning Representations}, 2021.

\bibitem[Li et~al.(2020)Li, Kovachki, Azizzadenesheli, Liu, Bhattacharya, Stuart, and Anandkumar]{Li2020MultipoleGNA}
Z.-Y. Li, N.~B. Kovachki, K.~Azizzadenesheli, B.~Liu, K.~Bhattacharya, A.~M. Stuart, and A.~Anandkumar.
\newblock Multipole graph neural operator for parametric partial differential equations.
\newblock \emph{ArXiv}, abs/2006.09535, 2020.
\newblock URL \url{https://arxiv.org/pdf/2006.09535.pdf}.

\bibitem[Long et~al.(2018)Long, Lu, Ma, and Dong]{long2018pde}
Z.~Long, Y.~Lu, X.~Ma, and B.~Dong.
\newblock Pde-net: Learning pdes from data.
\newblock In \emph{International conference on machine learning}, pages 3208--3216. PMLR, 2018.

\bibitem[Lu et~al.(2021)Lu, Jin, Pang, Zhang, and Karniadakis]{deeponet2021}
L.~Lu, P.~Jin, G.~Pang, Z.~Zhang, and G.~E. Karniadakis.
\newblock Learning nonlinear operators via deeponet based on the universal approximation theorem of operators.
\newblock \emph{Nature Machine Intelligence}, 3\penalty0 (3):\penalty0 218–229, Mar. 2021.
\newblock ISSN 2522-5839.
\newblock \doi{10.1038/s42256-021-00302-5}.
\newblock URL \url{http://dx.doi.org/10.1038/s42256-021-00302-5}.

\bibitem[Mao et~al.(2021)Mao, Lu, Marxen, Zaki, and Karniadakis]{mao2021deepm}
Z.~Mao, L.~Lu, O.~Marxen, T.~A. Zaki, and G.~E. Karniadakis.
\newblock Deepm\&mnet for hypersonics: Predicting the coupled flow and finite-rate chemistry behind a normal shock using neural-network approximation of operators.
\newblock \emph{Journal of computational physics}, 447:\penalty0 110698, 2021.

\bibitem[Molina et~al.(2019)Molina, Schramowski, and Kersting]{molina2019pad}
A.~Molina, P.~Schramowski, and K.~Kersting.
\newblock Pad$\backslash$'e activation units: End-to-end learning of flexible activation functions in deep networks.
\newblock \emph{arXiv preprint arXiv:1907.06732}, 2019.

\bibitem[Pathak et~al.(2022)Pathak, Subramanian, Harrington, Raja, Chattopadhyay, Mardani, Kurth, Hall, Li, Azizzadenesheli, et~al.]{pathak2022fourcastnet}
J.~Pathak, S.~Subramanian, P.~Harrington, S.~Raja, A.~Chattopadhyay, M.~Mardani, T.~Kurth, D.~Hall, Z.~Li, K.~Azizzadenesheli, et~al.
\newblock Fourcastnet: A global data-driven high-resolution weather model using adaptive fourier neural operators, arxiv [preprint], 10.48550.
\newblock \emph{arXiv preprint arXiv.2202.11214}, 22, 2022.

\bibitem[Raissi et~al.(2019)Raissi, Perdikaris, and Karniadakis]{raissi2019physics}
M.~Raissi, P.~Perdikaris, and G.~E. Karniadakis.
\newblock Physics-informed neural networks: A deep learning framework for solving forward and inverse problems involving nonlinear partial differential equations.
\newblock \emph{Journal of Computational physics}, 378:\penalty0 686--707, 2019.

\bibitem[Raoni'c et~al.(2023)Raoni'c, Molinaro, Ryck, Rohner, Bartolucci, Alaifari, Mishra, and de~B'ezenac]{Raonic2023ConvolutionalNOA}
B.~Raoni'c, R.~Molinaro, T.~D. Ryck, T.~Rohner, F.~Bartolucci, R.~Alaifari, S.~Mishra, and E.~de~B'ezenac.
\newblock Convolutional neural operators for robust and accurate learning of pdes.
\newblock In \emph{Neural Information Processing Systems}, 2023.
\newblock URL \url{https://api.semanticscholar.org/CorpusId:258968120}.

\bibitem[Ronneberger et~al.(2015)Ronneberger, Fischer, and Brox]{ronneberger2015u}
O.~Ronneberger, P.~Fischer, and T.~Brox.
\newblock U-net: Convolutional networks for biomedical image segmentation.
\newblock In \emph{Medical image computing and computer-assisted intervention--MICCAI 2015: 18th international conference, Munich, Germany, October 5-9, 2015, proceedings, part III 18}, pages 234--241. Springer, 2015.

\bibitem[Sau and Yin(2024)]{sau2024reviewneuralnetworksolvers}
R.~C. Sau and L.~Yin.
\newblock A review of neural network solvers for second-order boundary value problems, 2024.
\newblock URL \url{https://arxiv.org/abs/2407.00442}.

\bibitem[Shen et~al.(2024)Shen, Marwah, and Talwalkar]{shen2024ups}
J.~Shen, T.~Marwah, and A.~Talwalkar.
\newblock {UPS}: Efficiently building foundation models for {PDE} solving via cross-modal adaptation.
\newblock \emph{Transactions on Machine Learning Research}, 2024.
\newblock ISSN 2835-8856.
\newblock URL \url{https://openreview.net/forum?id=0r9mhjRv1E}.

\bibitem[Sushnikova et~al.(2022)Sushnikova, Kharyuk, and Oseledets]{sushnikova2022fmm}
D.~Sushnikova, P.~Kharyuk, and I.~Oseledets.
\newblock Fmm-net: neural network architecture based on the fast multipole method.
\newblock \emph{arXiv preprint arXiv:2212.12899}, 2022.

\bibitem[Telgarsky(2016)]{telgarsky2016benefits}
M.~Telgarsky.
\newblock Benefits of depth in neural networks.
\newblock In \emph{Conference on learning theory}, pages 1517--1539. PMLR, 2016.

\bibitem[Totounferoush et~al.(2025)Totounferoush, Kotchourko, Mahoney, and Staab]{totounferoush2025paving}
A.~Totounferoush, S.~Kotchourko, M.~W. Mahoney, and S.~Staab.
\newblock Paving the way for scientific foundation models: enhancing generalization and robustness in pdes with constraint-aware pre-training.
\newblock \emph{arXiv preprint arXiv:2503.19081}, 2025.

\bibitem[Vaswani et~al.(2017)Vaswani, Shazeer, Parmar, Uszkoreit, Jones, Gomez, Kaiser, and Polosukhin]{vaswani2017attention}
A.~Vaswani, N.~Shazeer, N.~Parmar, J.~Uszkoreit, L.~Jones, A.~N. Gomez, {\L}.~Kaiser, and I.~Polosukhin.
\newblock Attention is all you need.
\newblock \emph{Advances in neural information processing systems}, 30, 2017.

\bibitem[Wang et~al.(2024)Wang, Li, Dwivedi, Hara, and Wu]{beno2024}
H.~Wang, J.~Li, A.~Dwivedi, K.~Hara, and T.~Wu.
\newblock Beno: Boundary-embedded neural operators for elliptic pdes.
\newblock \emph{ArXiv}, abs/2401.09323, 2024.
\newblock URL \url{https://api.semanticscholar.org/CorpusId:267027614}.

\bibitem[Wang et~al.(2021)Wang, Wang, and Perdikaris]{wang2021learning}
S.~Wang, H.~Wang, and P.~Perdikaris.
\newblock Learning the solution operator of parametric partial differential equations with physics-informed deeponets.
\newblock \emph{Science advances}, 7\penalty0 (40):\penalty0 eabi8605, 2021.

\bibitem[Wu et~al.(2024)Wu, Luo, Wang, Wang, and Long]{Wu2024TransolverAFA}
H.~Wu, H.~Luo, H.~Wang, J.~Wang, and M.~Long.
\newblock Transolver: A fast transformer solver for pdes on general geometries.
\newblock \emph{ArXiv}, abs/2402.02366, 2024.
\newblock URL \url{https://api.semanticscholar.org/CorpusId:267411758}.

\bibitem[Xia et~al.(2010)Xia, Chandrasekaran, Gu, and Li]{xia2010fast}
J.~Xia, S.~Chandrasekaran, M.~Gu, and X.~S. Li.
\newblock Fast algorithms for hierarchically semiseparable matrices.
\newblock \emph{Numerical Linear Algebra with Applications}, 17\penalty0 (6):\penalty0 953--976, 2010.

\bibitem[Zhang et~al.(2024)Zhang, Kahana, Kopani{\v{c}}{\'a}kov{\'a}, Turkel, Ranade, Pathak, and Karniadakis]{zhang2024blending}
E.~Zhang, A.~Kahana, A.~Kopani{\v{c}}{\'a}kov{\'a}, E.~Turkel, R.~Ranade, J.~Pathak, and G.~E. Karniadakis.
\newblock Blending neural operators and relaxation methods in pde numerical solvers.
\newblock \emph{Nature Machine Intelligence}, pages 1--11, 2024.

\end{thebibliography}
}


\appendix


\section{FMM Matrix-vector product}
\begin{figure}
    \centering
    \includegraphics[width=0.7\linewidth]{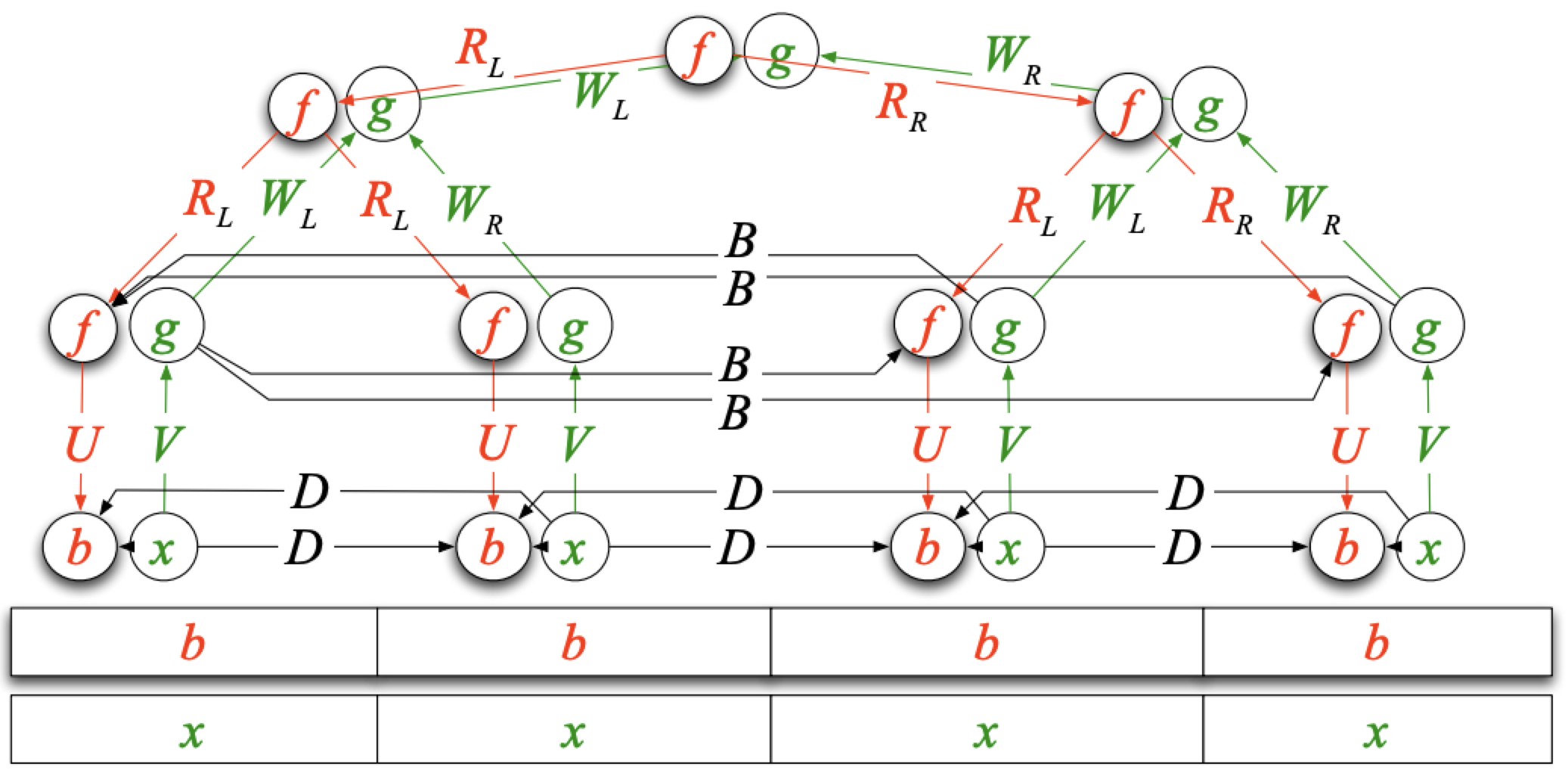}
    \caption{Computational graph of FMM matrix-vector product. }
    \label{fig:fmm_sfg}
\end{figure}

Figure \ref{fig:fmm_sfg} shows the computational graph of a matrix-vector product $Ax=b$ using fast multiple method (FMM)\citep{greengard1987fast}. The graph shows the input block vector $x$ being transformed through up-sweep and down-sweep recursions as:
\begin{align*}
    g_i &= V_ix_i + \sum_{j\in C(i)}W_j g_j; \\
    f_i &= R_{C^{-1}(i)} f_{C^{-1}(i)} + \sum_{j\in S(i)} B_{i,j}g_j; \\
    b_i &= D_i x_i + U_i f_i;
\end{align*}
where $C(i)$ denote the set of child nodes of node $i$, $C^{-1}(i)$ denote the parent nodes of node $i$, and $S(i)$ denote the set of neighbors of node $i$, such that the corresponding edges are not in the tree graph. The indices are omitted from the figure for brevity. The edges $B,D$ correspond to bridge operators and green and red paths correspond to encoder and decoder of Figure \ref{fig:fmm_block}. The FMM recursions are sparse linear equations in $x,b,f,g$ and so the matrix-vector product is computed efficiently.

\section{Full architecture of MNO}
\begin{figure}
    \centering
    \includegraphics[width=\linewidth]{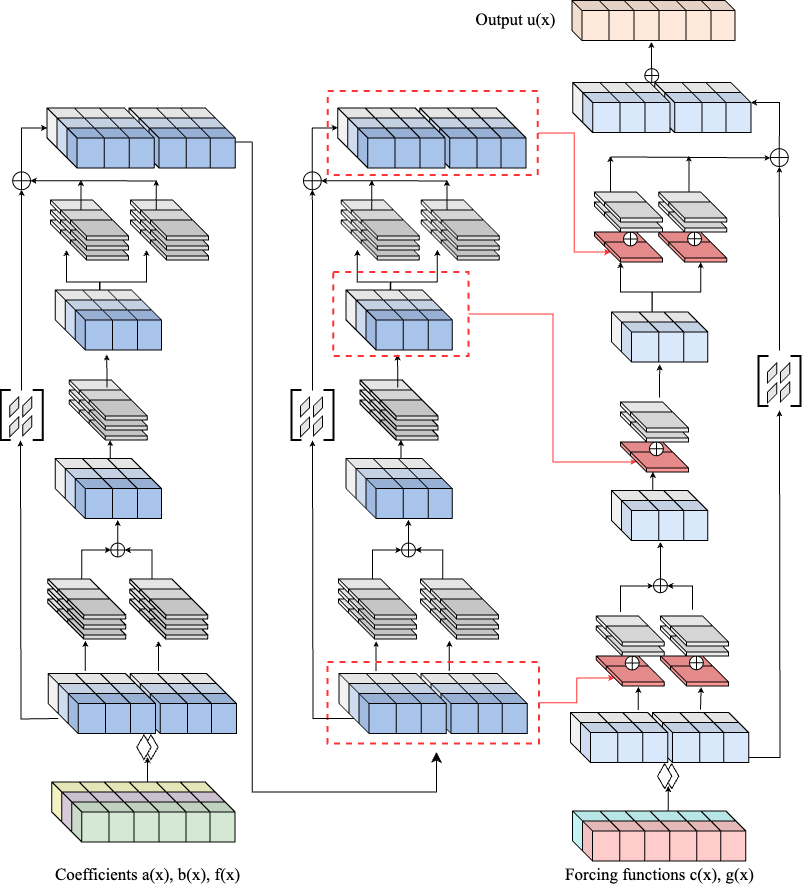}
    \caption{MNO with two GFMM-blocks in the coefficient branch and one GFMM-block in the RHS-branch showing multi-modal fusion between the two branches.}
    \label{fig:mlnet}
\end{figure}

The full architecture of MNO for multi-channel inputs and multi-modal fusion between coefficient and RHS-branches is shown in Figure \ref{fig:mlnet}. The illustration shows two GFMM-blocks in the coefficient branch and one GFMM-block in the RHS-branch. Considering the example of multi-parameter BVP of (\ref{eq:integralequation}), the coefficients of interior and boundary equations, $a(x), b(x) \text{ and } f(x)$ vectors are stacked and given as input to the coefficient branch. RHS terms $c(x), g(x)$ are given as input to the RHS-branch. The weights of the RHS-branch are corrected additively with the intermediate outputs from the coefficient branch and the entire network is trained end-to-end. At each encoder and decoder layer of the GFMM-blocks, the input is transformed using the \texttt{BasisTransform} described in algorithm \ref{alg:multi_domain} which also includes the weight correction from multi-modal fusion for the GFMM-blocks in RHS-branch.

\begin{algorithm}
\caption{Basis Transformation inside GFMM-block}\label{alg:multi_domain}
\begin{algorithmic}[1]
\Input $\mathbf{F} \in \mathbb{R}^{C_{out}\times C_{in} \times P\times P}$: Encoder/decoder parameters
\Input $y \in \mathbb{R}^{C_{in} \times P}$
\Input $\varepsilon \in \mathbb{R}^{P \times P}$ \Comment{Latent outputs (for GFMM-blocks with multi-modal fusion)}
\Output $z \in \mathbb{R}^{C_{out} \times P}$
\Parameters
    \State $C_{in}$: Number of input channels
    \State $C_{out}$: Number of output channels
    \State $\phi: \mathbb{R}^d \rightarrow \mathbb{R}^d$: Element-wise nonlinear activation function
    \State $\mathcal{E}$: True if this is a GFMM-block with multi-modal fusion.
\Procedure{BasisTransform}{$\mathbf{F}, y, \varepsilon$}
\State $z = \mathbf{0}$
\For{$i=1$ to $C_{out}$}
    \For{$j=1$ to $C_{in}$}
        \If { $\mathcal{E}$} 
            \State $\mathbf{F}_{ij} \gets \mathbf{F}_{ij} + \varepsilon$ \Comment{Additive weight correction (multi-modal fusion)}
        \EndIf
        \State $z_{i} \gets z_{i} + \mathbf{F}_{ij} y_j$ \Comment{Matrix-vector product and sum over input channels}
    \EndFor
\EndFor
\State $z_i \gets \phi(z_i)$ for $i=1$ to $C_{out}$ \Comment{nonlinear activation}
\EndProcedure
\end{algorithmic}
\end{algorithm}

\begin{figure}[ht]
    \centering
    \includegraphics[width=0.5\linewidth]{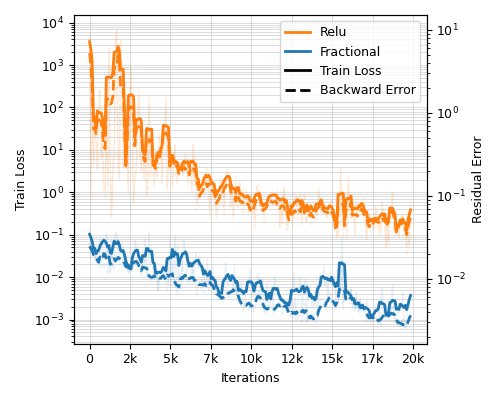}
    \caption{Training MSE loss and residual error of 1D Darcy flow MNO network with relu vs nonlinear rational activation function.}
    \label{fig:relu_vs_frac}
\end{figure}

\section{ReLU vs Rational Activation} \label{sec:relu_vs_rational}

We make the observation that the solution to a linear partial differential equation is a rational function of its coefficients. Moreover, to achieve the multi-modal fusion, we produce the weight corrections of the RHS-branch with the intermediate basis representations of the coefficient branch. Motivated by these, we explore the performance of rational functions as activations for our neural operator networks, in contrast to the traditional ReLU. We empirically observe that a non-linear rational function such as $\phi (x) = \frac{x}{1+|x|}$ performs better than ReLU for learning the multi-modal solution operators of both linear and non-linear BVPs. Figure \ref{fig:relu_vs_frac} shows the training loss and residual error progression of MNO models on 1D Darcy flow and the nonlinear rational function outperforms ReLU by a big margin. Previous works~\citep{molina2019pad} explored the variations of learnable rational functions as activations in neural networks. A rigorous analysis of choice of activation functions for MNOs is planned in our future work.

\begin{table}[ht]
    \small
    \centering
    {\setlength{\tabcolsep}{3pt}
    {
     \caption{Same as Table \ref{tab:p1d_results} but with mean and standard deviations reported from 3 independent training runs of each model.}
     \label{tab:p1d_results_stats}
     
     \begin{subtable}{\textwidth} 
         \centering
         \caption{Solution Sampled}
         \label{tab:p1d_results_stats_sol}
         \begin{tabular}{lcccc}
             \toprule
             \multirow{2}{*}{Model} & \multicolumn{2}{c}{Chebyshev} & \multicolumn{2}{c}{Mixed} \\
             \cmidrule(lr){2-3} \cmidrule(lr){4-5}
             & $\epsilon_{\text{rel}}$ & $\epsilon_{\text{be}}$ & $\epsilon_{\text{rel}}$ & $\epsilon_{\text{be}}$ \\
             \midrule
             FNO & (8.24$\pm$0.03)e-03 & (1.72$\pm$0.05)e-3 & (4.54$\pm$0.07)e-01 & (3.29$\pm$0.02)e-2 \\
             DeepONet & (2.05$\pm$0.06)e-03 & (1.60$\pm$0.09)e-3 & (1.09$\pm$0.02)e-03 & (2.40$\pm$0.03)e-3 \\
             UNO & \textbf{(5.32$\pm$0.01)e-06} & \textbf{(1.12$\pm$0.02)e-4} & \textbf{(5.72$\pm$0.03)e-06} & \textbf{(1.14$\pm$0.05)e-4} \\
             \bottomrule
         \end{tabular}
     \end{subtable}
     
     \vspace{1em} 
     
     \begin{subtable}{\textwidth} 
         \centering
         \caption{RHS Sampled}
         \label{tab:p1d_results_stats_rhs}
         \begin{tabular}{lcccc}
             \toprule
             \multirow{2}{*}{Model} & \multicolumn{2}{c}{Chebyshev} & \multicolumn{2}{c}{Mixed} \\
             \cmidrule(lr){2-3} \cmidrule(lr){4-5}
             & $\epsilon_{\text{rel}}$ & $\epsilon_{\text{be}}$ & $\epsilon_{\text{rel}}$ & $\epsilon_{\text{be}}$ \\
             \midrule
             FNO & (7.56$\pm$0.08)e-01 & (1.29$\pm$0.04)e-2 & (9.63$\pm$0.01)e-01 & (2.15$\pm$0.06)e-2 \\
             DeepONet & (2.63$\pm$0.05)e-01 & (1.98$\pm$0.07)e-2 & (7.31$\pm$0.04)e-01 & (1.45$\pm$0.08)e-2 \\
             UNO & \textbf{(8.18$\pm$0.07)e-02} & \textbf{(2.08$\pm$0.04)e-3} & \textbf{(5.76$\pm$0.09)e-01} & \textbf{(4.78$\pm$0.02)e-3} \\
             \bottomrule
         \end{tabular}
     \end{subtable}
     
     }} 
\end{table}

\section{1D Poisson}
\subsection{Results from multiple training runs} \label{sec:p1d_stats}
We run the experiment outlined in section \ref{sec:p1d} by training the FNO, DeepONet and UNO models independently three times and aggregate the results in Table \ref{tab:p1d_results_stats} which reports the mean and standard deviation of the relative and residual errors for the three models in the 1D Poisson case.
\subsection{Testing on out-of-training distribution}
\begin{figure}[ht]
    \centering
    \includegraphics[width=0.8\linewidth]{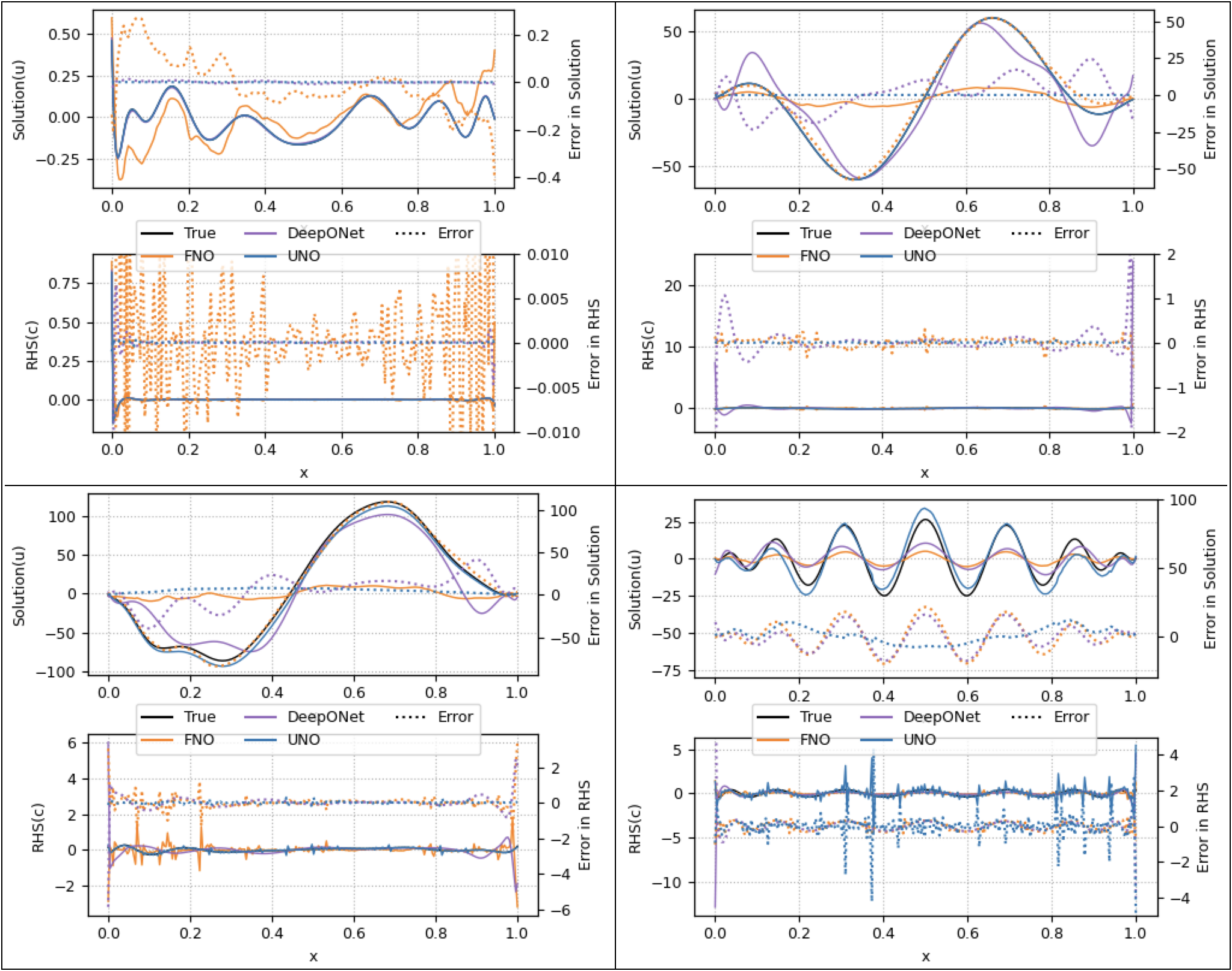}
    \caption{Examples of UNO, FNO and DeepONet evaluated on out-of-distribution 1D Poisson data. Top left sample is generated using solution sampling with random linear combinations of Chebyshev bases. Rest of the plots show samples generated using RHS sampling.}
    \label{fig:p1d_mixcheb1}
\end{figure}

Figure~\ref{fig:p1d_mixcheb1} shows the UNO, FNO, DeepONet models which are trained using solution sampling using randomly scaled Chebyshev polynomials i.e. $u = \alpha_k T_k(x)$ from \ref{sec:p1d}, when they are tested on out-of-training distribution examples.  For the top left plot, the example is generated by solution sampling, but groundtruth $u$ is chosen as random linear combination of $T_1(x)$ to $T_{16}(x)$. The figure shows that UNO and DeepONet performs better than FNO on this type of examples. In the rest of the plots, examples are generated using RHS sampling. For the top right and bottom right plots, $c$ is chosen as randomly scaled $T_k(x)$ where $k=4$ for top right and $k=16$ for bottom right plot and groundtruth $u$ is correspondingly derived. For the bottom left plot, $c$ is chosen as a random linear combination from $T_1(x)$ to $T_{16}(x)$. For all the RHS-sampled examples, every model struggles to approximate the solution, but UNO performs better with lower residual error compared to FNO and DeepONet.

\section{2D GFMM-block} \label{sec:2d}
\subsection{Architecture}

\begin{figure}[ht]
    \centering
    \includegraphics[width=\linewidth]{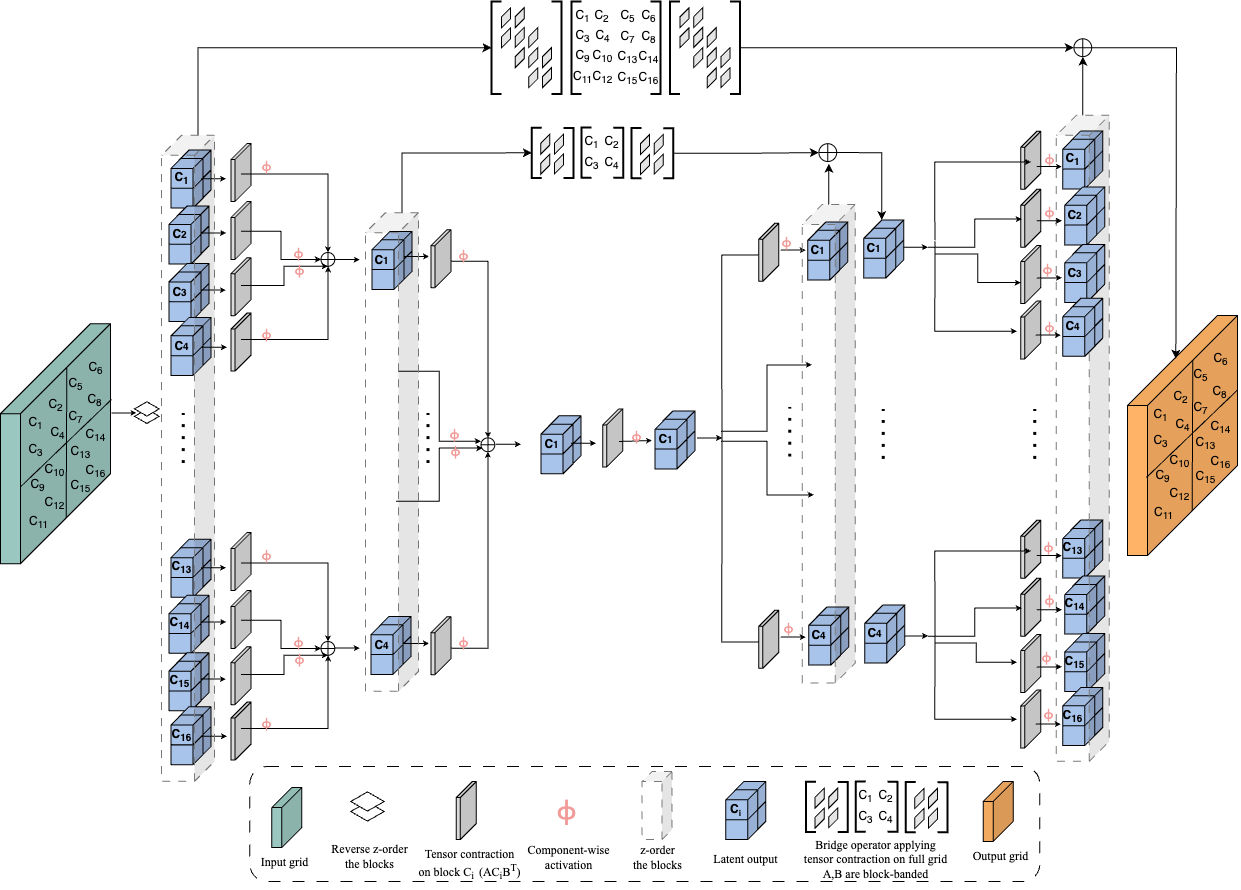}
    \caption{Schematic of a 2D GFMM block.}
    \label{fig:fmm2d}
\end{figure}

A 2D FMM computational graph is equivalent to the tensor product of a 1D FMM shown in Figure~\ref{fig:fmm_sfg}. Analogously, we define the 2D GFMM-block as the tensor-product of 1D GFMM-block. A schematic of the 2D GFMM-block architecture is shown in Figure~\ref{fig:fmm2d}. For the simplest case of single channel 2D input grid, the encoder and decoder graphs are quadtrees as compared to the binary trees of the 1D FMM. Each encoder node comprises of four bi-linear transformation layers that apply a \emph{tensor contraction} operation on the input blocks and add the result into the output block. Tensor contraction is defined as $A X B^T$, where $A$ and $B$ are the linear weights and $X$ is the input block.

The first encoder layer's input blocks are block matrices of the input 2D grid array in a fixed order. In Figure~\ref{fig:fmm2d}, we used Z-ordering (or Morton ordering) to divide the 2D input matrix into blocks. The rest of the encoder nodes follow the similar quadtree structure shown. Similarly, at every decoder layer, each input block undergoes four bi-linear transformations (tensor contraction) and the output is added to the output of bridge operator, which performs the tensor contraction on the corresponding encoder output. Finally, the output is obtained by adding the output of last decoder layer with the transformed input through the outermost bridge operator. For the bridge operators, the linear weights are fixed to be block banded matrices (block tri-diagonal in our experiments). Similar to 1D GFMM-block, we only store the blocks on the banded diagonals for memory efficiency. Additional non-linear activations can be applied after each bi-linear transformation of the encoder and decoder layers. In the next section, we show empirically that a 2D GFMM-block thus constructed can approximate the inverse operator of a discrete 2D Poisson's equation.

\subsection{2D Discrete Poisson}
\begin{figure}
    \centering
    \includegraphics[width=\linewidth]{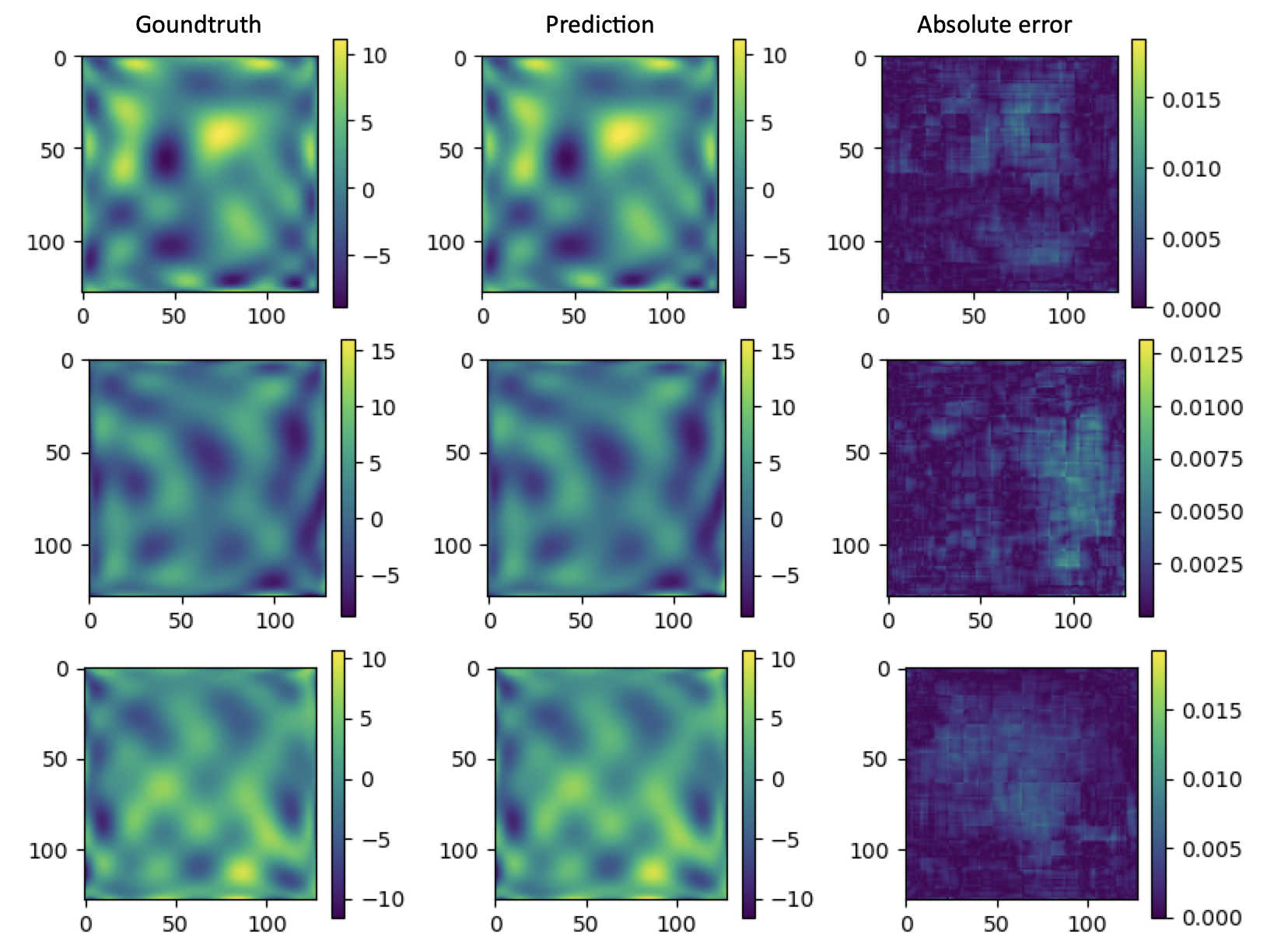}
    \caption{Groundtruth solution and the predicted output of the 2D GFMM-block trained on 2D Poisson's equation. }
    \label{fig:p2d}
\end{figure}

Similar to section \ref{sec:p1d}, we consider the linear equation $Au=c$, but here $A$ is the 2D discrete Laplacian operator. We train the 2D GFMM-block to learn the inverse operator of $A$ using solution sampling of random linear combinations of Chebyshev polynomials on a $128\times 128$ uniform grid. We initialize the 2D GFMM-block with four encoder and decoder layers and a block size of 8 and trained for 200k iterations. Figure~\ref{fig:p2d} shows the results on randomly sampled examples from 2D Chebyshev grid where the 2D GFMM-block network is able to approximate the inverse operator on this simple setup. It obtains a mean relative error of (4.86$\pm$0.3)e-07 and a normalized backward error of (6.48$\pm$0.005)e-05 on a random test set of 1000 examples.

Given the encouraging results, extending the 2D GFMM-block architecture with multi-modal fusion and a careful analysis of its generalization performance is the subject of our future work.

\end{document}